# High Entropy Oxides as Room Temperature Anderson Insulators with Colossal Soft Gaps

*Elena Salagre[1], Heather Hare[2], Eli Rotenberg[3], Jonathan D. Denlinger [3], Joshua D. Sugar[1], Yiyang Li[2], and Elliot J. Fuller[1]**

[1]Sandia National Laboratories, 7011 East Ave. Livermore, CA 94550, United States of America
[2] Materials Science and Engineering, University of Michigan, Ann Arbor, MI, 48109, United States of America
[3]Advanced Light Source, Lawrence Berkeley National Laboratory, Berkeley, California 94720, United States of America

*Correspondence: ejfull@sandia.gov

## ABSTRACT

The Anderson effect, arising from destructive quantum interference of electron wave packets in disordered potentials, is usually observed at cryogenic temperatures and typically arises from extrinsic defects in conventional materials such as phosphorus dopants in silicon or oxygen vacancies in transition metal oxides. However, here we demonstrate that a multi-cation lattice of transition metals in a high entropy oxide (HEO) natively establishes an interacting Anderson insulator that extends up to 500 K. By controlling the population of transition metal cations in $\mathrm{La_{0.7}Sr_{0.3}Mn_{1-x}(Cr,Fe,Co,Ni)_{x/4}O_3}$ (LS5B), we tune the system from the prototypical correlated metal $\mathrm{La_{0.7}Sr_{0.3}MnO_3}$ at ($x = 0.00$) to a localized insulator ($x = 0.80$), where the disorder length $\xi_{dis}$ contracts to near the unit cell (3.85 Å) which drives the electron mean free path to the Ioffe-Regel limit. Angle-resolved photoemission spectroscopy (ARPES) reveals the systematic localization of itinerant $e_g$ quasiparticle states at the Fermi level, consistent with the emergence of a soft Coulomb gap, while the underlying band structure and symmetry are preserved. Transport reveals Efros-Shklovskii variable-range hopping governed by a colossal soft gap with characteristic temperature $T_{ES}$ > 50,000 K and screened Coulomb interaction strengths as high as 1.6 eV. Together, multiple lines of evidence indicate interacting Anderson behavior as a native characteristic that survives to temperatures as high as 500 K in high entropy materials.

**INTRODUCTION**

Anderson insulators are observed in a broad array of physical systems including heavily doped semiconductors, transition metal oxides and in diffuse scattering media in the optical domain[1–5]. However, the Anderson effect typically arises from extrinsic material effects, such as defects or grain boundaries. Therefore, quantum mechanical destructive interference has remained largely an edge case, for example, after heavy doping beyond practical application or limited to cryogenic temperatures. The effect is not considered to be the native property of a material class in ambient conditions. However, the emerging class of high entropy, or compositionally complex materials (CCM) sits at an extreme condition which challenges this notion by blending what is considered a native property versus a defect. In CCM share multiple elements (> 4) to a lattice site but can support near solid-solutions with strong site potential variations $V_i$ and a nominal site substitution spacing $\xi_{dis}$ that approaches the unit cell dimensions[6–10]. These criteria meet conditions for Anderson localization where the electronic mean free path $l$ reaches the Ioffe-Regel limit $k_F l \sim 1$, where $k_F$ is the Fermi momentum[11]. Within CCM, compositionally complex oxides, herein referred to as HEO, present an intriguing case where an oxygen sublattice often enforces a highly ordered and crystalline material in addition to high intermixing of shared cations. While HEO are known to host large chemical disorder, it has not been directly established whether this disorder produces an Anderson-localized ground state rather than conventional polaronic, Mott, granular, or defect-mediated insulating behavior. Furthermore, given the relevance of HEOs as photocatalysts, battery particles and energy storage ceramics, a framework for understanding their electronic behavior is critical to their widespread technological impact, which we discuss later[12–15].

Here, we establish the compound LS5B as supporting native Anderson behavior that extends up to 500 K. To achieve this, we introduced transition metal complexity into the prototypical perovskite oxide $La_{0.7}Sr_{0.3}MnO_3$ (LSMO), depicted in the schematic in Figure 1a. LSMO is a well understood double-exchange metal with clearly dispersive and itinerant $e_g$ hole carriers that offer an ideal ground state. We gradually substituted Mn with an equimolar blend of additional transition metal cations (Cr, Fe, Co and Ni) to observe LSMO evolve into a chemically complex, high-entropy electronic system $\mathrm{La_{0.7}Sr_{0.3}Mn_{1-x}(Cr,Fe,Co,Ni)_{x/4}O_3}$ (LS5B). Through gradual substitution, our approach enabled us to quantify electronic properties as a function of a calculated disorder spacing $\xi_{dis} = \frac{a}{x^{1/3}}$ plotted in Figure 1b, by approximating a solid solution intermixing of B-site cations according to $x$ and using the lattice parameter $a$ from LSMO. This length extends from theoretically infinite dimensions for a perfect LSMO crystal at $x = 0.00$, to roughly 1 nm at $x = 0.10$, and down to a single unit cell $a \approx 3.85$ Å at $x = 0.80$. As $x$ increases, transition metal site potential variations $V_i(r)$ localize carriers as depicted in the Figure 1c schematic. Here, we plot the carrier wavefunction envelope $F(r)$ from $\psi(r) = F(r)u(r)$, where $\psi(r)$ is the wavefunction and $u(r)$ is the lattice-periodic part of the Bloch function. For a theoretically perfect LSMO lattice, the envelope is flat, representing itinerant, delocalized carriers. However, when a critical disorder length is reached, the wavefunction becomes strongly localized and transport is governed by variable range hopping (Fig. 1c, last panel). An interacting Anderson insulator is a combination of the strong localization due to varying site potential with the addition of Coulomb interactions[16–19]. In this work, multiple independent signatures including Efros–Shklovskii variable range hopping, a collapse of ARPES quasiparticle coherence toward the Ioffe–Regel limit, and epitaxial microstructure without extended defects indicate a native interacting Anderson state.

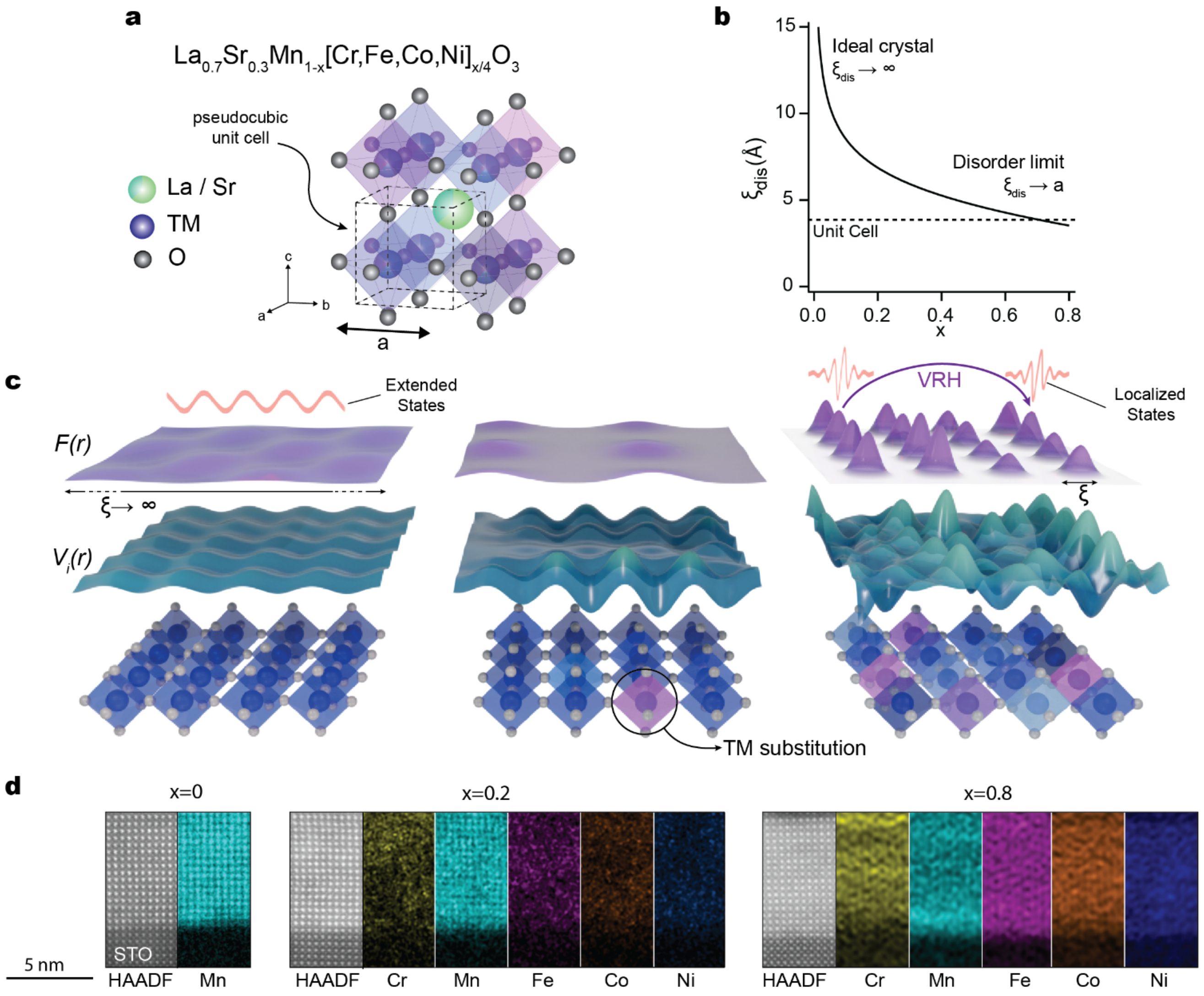

**Figure 1**. **(a)** Pseudocubic representation of the LS5B $R\bar{3}c$ structure. **(b)** Calculated disorder spacing based on substitution per unit cell. **(c)** Schematic of substitution effects in the periodic crystal lattice, where blue denotes the parent compound (LSMO) Mn-O octahedra and purple, pink and cyan sites denote the substituted TM-O. On top of the crystal, the single site potential for a simplified cubic lattice is plotted. Above this, the wave function envelope $F(r)$ is depicted in purple to highlight the decrease in localization length and the confinement of carriers. **(d)** HAADF sections and standardless-quantified EDS composition maps of LSMO and substituted compounds from left to right x=0, 0.2 and 0.8. Colors represent estimated elemental atomic fractions. The Mn maps use a display range of 0–0.30 for x = 0 and 0.2 to avoid saturation. For x = 0.8, all elemental maps, including Mn, use the same display range of 0–0.10. Color brightness should therefore be compared only between maps with the same display range.

## RESULTS

We grew ~9 nm thin films onto $SrTiO_3$ single crystal substrates with a (100) orientation using pulsed laser deposition. To eliminate the effects of extended defects such as grain boundaries, the samples were grown as epitaxial films at 650 °C. The substrate $SrTiO_3$ was used to maintain a small $1 - 2\%$ in-plane tensile strain across the entire series from $x = 0.00 - 0.80$ (see Supplementary Text S2), in agreement with previous work[8] . High oxygen partial pressure during growth (300 mTorr) and post growth annealing in atmosphere were used to oxygenate the films and reduce the effect of oxygen vacancies. Samples were then characterized by X-ray diffraction (XRD), X-ray photoelectron spectroscopy (XPS) and high-angle annular dark-field scanning transmission electron microscopy (HAADF-STEM). The representative regions shown in Figure 1d exhibit substantial transition-metal intermixing. While EDS through-thickness profiles reveal minor surface/interface compositional variations (such as Sr and Cr enrichment; see Supplementary Section S7), the HAADF STEM images show no resolved grain boundaries, dislocations or lateral segregated elemental domains.

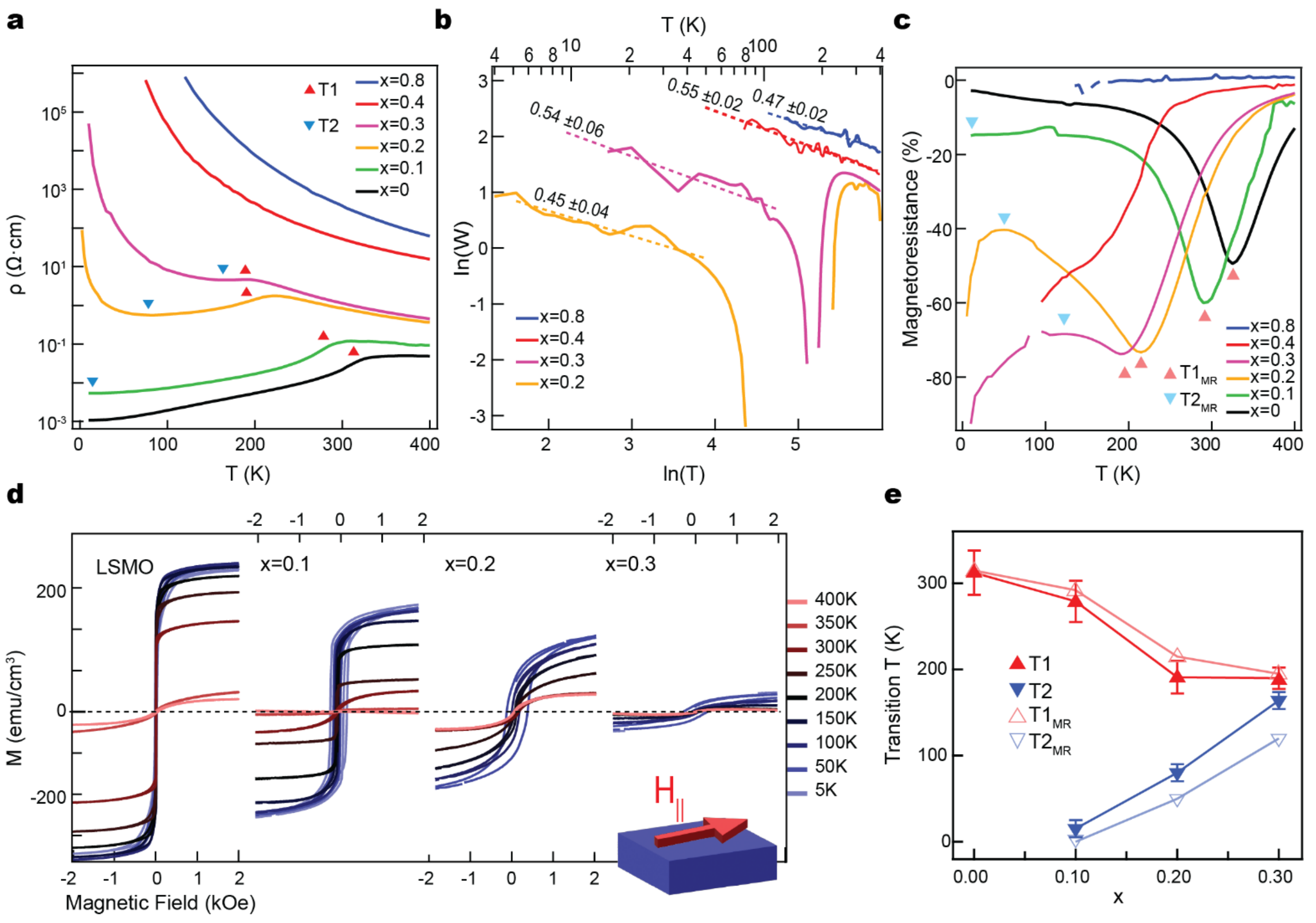

**Figure 2. (a)** Resistivity for ~9 nm thin films between 5 and 400 K. **(b)** Zabrodskii plot and exponent analysis of electronic transport. The generalized exponents have been extracted using the analysis described in the main text and SI and shown next to each linear fit. **(c)** Magnetoresistance percentage curves for H=0 - 9T (90 kOe). **(d)** Magnetization loops as a function of applied field and temperature for 50 nm films. Field was applied in-plane. **(e)** Temperature evolution of the metal insulator transitions and magnetoresistance inflection points**.** In (a) and (c) transition temperatures are shown by colored triangles.

As plotted in Figure 2a, the LS5B system shows distinct evolution of resistivity with temperature. For $x = 0.00$ (black line), the LSMO ferromagnetic metallic phase (FMM) is observed at low temperature with a well-known transition to a paramagnetic insulating phase (PMI) at around 350 K[20]. Over the range $x = 0.10 - 0.30$ (green, orange and pink lines) this transition T1 (identified with red triangles) decreases monotonically from 310 K to 190 K. Interestingly, a new insulating transition (T2 blue triangles) emerges at low temperature that is not present in LSMO, and that exists even for dilute substitutions ($x = 0.10$). We associate this new transition with a new insulating phase (I2) and find that T2 increases in temperature from about 80 K at $x = 0.20$ until 160 K at $x = 0.30$. At $x = 0.10$, the T2 upturn occurs at 15 K and the I2 phase occupies only a very small temperature window. For samples with $x \geq 0.40$, I2 exists for all temperatures measured. PMI behavior is not observed and therefore T2 likely sits at a temperature above the measured range (see Supplementary Text S3).

To describe the observed transport and identify the governing mechanisms, we use the generalized transport equation:

$$\rho(T) = \rho_0 \exp\left[\left(\frac{T_0}{T}\right)^p\right] \qquad (1)$$

where $\rho(T)$ is the temperature dependent resistivity, $\rho_0$ is the resistivity prefactor, $T_0$ is the characteristic temperature and $p$ is the generalized transport exponent. The Zabrodskii method of reduced activation energy $W(T) = -\frac{\partial ln\rho}{\partial lnT}$ was used to determine the exponent $p$ in Eq. 1 which indicates different types of conduction[21,22]. For the Zabrodskii plot in Figure 2b, the data were fitted to a line over regions between the transitions (see S3 for details). For the high temperature PMI phase I1, the exponents ranged from $p = 0.96 - 1.10$ for $x = 0.00 - 0.30$, which is consistent with small polaron hopping and is expected for the PMI in LSMO[23–27]. For the I2 phase, we find exponents ranging from $p = 0.45 - 0.55$ over the composition $x = 0.20 - 0.80$. Exponents close to $p = 1/2$ are consistent with Efros-Shklovskii Variable Range Hopping (ES-VRH)[28,29]. The $p = 1/2$ exponent, together with the absence of resolved extended defects or grain boundaries in the HAADF-STEM images in Fig. 1d, is a hallmark characteristic of an interacting localized regime. Surprisingly, for $x \geq 0.40$ the slope of $p \approx 1/2$ persisted up to 500 K (for high temperature regime, see Supplementary text S3). Small deviations at the highest temperatures for the case of $x = 0.40$ suggest a potential crossover regime to a mix between an Anderson soft gap and a hard polaronic gap, but $x = 0.80$ maintains a slope of $p \approx 1/2$ for all temperatures measured.

Magnetoresistance (MR) and magnetization measurements (Figure 2c, d) complement our electronic transport data, with MR inflection points matching the T1 and T2 transitions. In the x = 0.00 (LSMO) baseline, the classic ~50% colossal magnetoresistance dip at 315 K aligns with the FMM-PMI transition (T1), indicating clean double-exchange transport[30,31]. As compositional complexity increases (x = 0.10 to 0.30), this T1 feature shifts to lower temperatures, and the MR magnitude increases up to 70%. Increased MR has been previously reported for LSMO derived compounds but not yet for compositionally complex counterparts[30,31]. Simultaneously, a secondary MR inflection point emerges at T2 for $x = 0.20, 0.30$, developing a negative MR slope. This I2 feature could reflect the field sensitivity of localized hopping transport, encompassing spin alignment, cluster polarization, and field-induced suppression of quantum-interference corrections. The excellent agreement between transport and MR transition temperatures (Figure 2e) confirms a composition-driven crossover from the FMM-PMI boundary (T1) to a localization-driven regime (T2). Magnetization loops (Fig. 2d) corroborate the fragmentation of the double-exchange network. While LSMO exhibits strong low-temperature ferromagnetism, the ferromagnetic character becomes extremely weak by x = 0.30, dropping below 50 $\mathrm{emu/cm^3}$ at larger $x$. Recent depth-resolved muon spectroscopy of Sr-free B-site-complex perovskite films revealed that chemical disorder in the transition metal sublattice generates a low-temperature crossover consistent with frustrated magnetism in a strongly correlated spin lattice[32]. In close agreement with these observations, our macroscopic magnetization data confirm that as complexity increases in LS5B, isolated, frustrated ferromagnetic regions become fully embedded within a predominantly localized background.

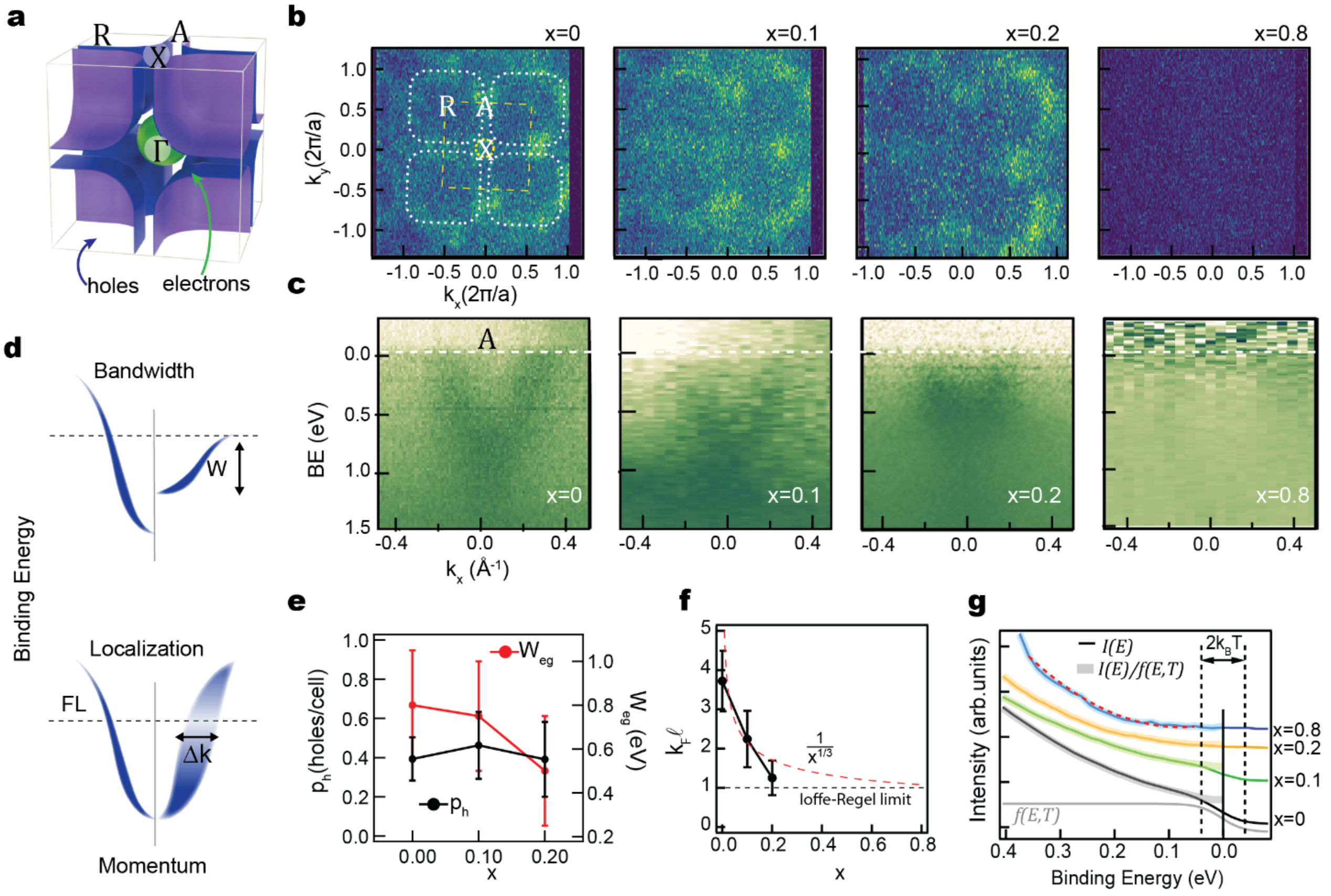

**Figure 3.** **(a)** 3D representation of canonical LSMO Fermi surface. Hole pockets shown in purple and electron pocket shown in green. **(b)** Fermi surfaces as a function of composition. The hole and electron pockets as well as relevant symmetry points are highlighted for x=0. **(c)** Normalized zoomed region of the $e_g$ states, specifically the hole pocket, with the $k_x$ dispersion centered at A symmetry point **(d)** Schematics of possible valence band effects based on bandwidth changes and localization. **(e)** Calculated carrier density from Luttinger theorem using approximately spherical 3D cube-like pockets extracted from (b) and ARPES derived occupied bandwidth of the $e_g$ band ($W_{eg}$). **(f)** Extracted single particle correlation parameter ($k_F/\Delta k$) showing Ioffe-Regel limit in dotted black line and a unit cell disorder length scale in dotted red line. **(g)** Integrated valence band for x=0, 0.1, 0.2 and 0.8 showcasing the intensity loss at $E_F$. Data corrected by the Fermi Dirac function is shown in lighter colors. Red dotted line shows a fit for the expected soft gap region in the x=0.8 case. Data for metallic regimes were acquired at 100K; data for x=0.8 was acquired at 200 K to avoid charging effects. Horizontal linear polarization and hv=131 eV was used to explore the hole pocket feature.

Angle resolved photoemission spectroscopy (ARPES) was used to directly measure the electronic band structure of the samples and to reveal the signatures of an interacting Anderson insulator. We begin with the ordered ground state of LSMO, with its Fermi surface shown in Figure 3a, and which is composed of Mn 3d $e_g$ orbitals. An electron pocket is centered at the Γ point and four hole-pockets (which dominate conduction) are centered at the corners of the Brillouin zone at the R symmetry points[26,33]. In Figure 3b the Fermi surfaces of samples $x = 0.00, 0.10, 0.20$ conserve the four hole-pocket symmetry. At $x = 0.80$, the sample is completely insulating at all temperatures, and the Fermi surface shows only very weak diffuse intensity. However, while the coherent $e_g$ quasiparticle weight vanishes, the deeper non-bonding $t_{2g}$ and O $2p$ bands retain dispersive character similar of LSMO, albeit with interaction driven broadening (See Supplementary Section S6). The band dispersions of the hole pockets are plotted in Figure 3c. For LSMO, highly dispersive bands are observed extending from ~1.5 eV to the Fermi level, indicating metallic conduction in close agreement with previous work[26,34,35]. For samples $x = 0.00 - 0.80$, the spectral weight near the Fermi level systematically grows more diffuse with larger $x$. Further, the overall band width is reduced as the tail of the dispersive feature moves toward the Fermi level while the minimum broadens in energy. The broadening suggests electron-electron interactions and shorter quasiparticle lifetimes.

To further establish the effects of Anderson localization, we performed hole-density and Ioffe-Regel analysis, as plotted in Fig. 3e and 3f respectively. For conductive samples that allowed analysis of the Fermi surface ($x = 0.00 - 0.20$), we found only minor differences in hole concentration, as extracted from momentum distribution curves (MDCs). This suggests that the main effect is localization rather than doping or band gap effects. We performed an Ioffe-Regel analysis in Figure 3f by approximating $k_F \ell \sim \frac{k_F}{\Delta k}$ where $\Delta k$ is the momentum width of the hole band crossing at the Fermi level (See Supplementary Text S6). From this calculation, we find that the system approaches the localization limit $k_F \ell \sim 1$ with increasing $x$ in qualitative agreement with our calculations of a disorder length based upon composition (Fig. 1b). For $x = 0.80$, a depletion in the photoemission intensity around the Fermi level is identified as a reduction of $e_g$ derived density. For a three dimensional interacting Anderson insulator, we would expect the density of states (DOS) to have the form $g(E) \propto |E - E_F|^2$ with finite spectral weight above the $t_{2g}$ band[28]. For the case of $x = 0.80$ the data was fit to this quadratic function (Figure 3g, red dashed line) and found to span a region with a width of at least ~350 meV consistent with a soft Coulomb gap. More details of the fitting can be found in Supplementary S6. In summary, the ARPES data support the picture of an Anderson insulator with the emergence of a soft Coulomb gap and closely agree with our electronic and magnetic measurements.

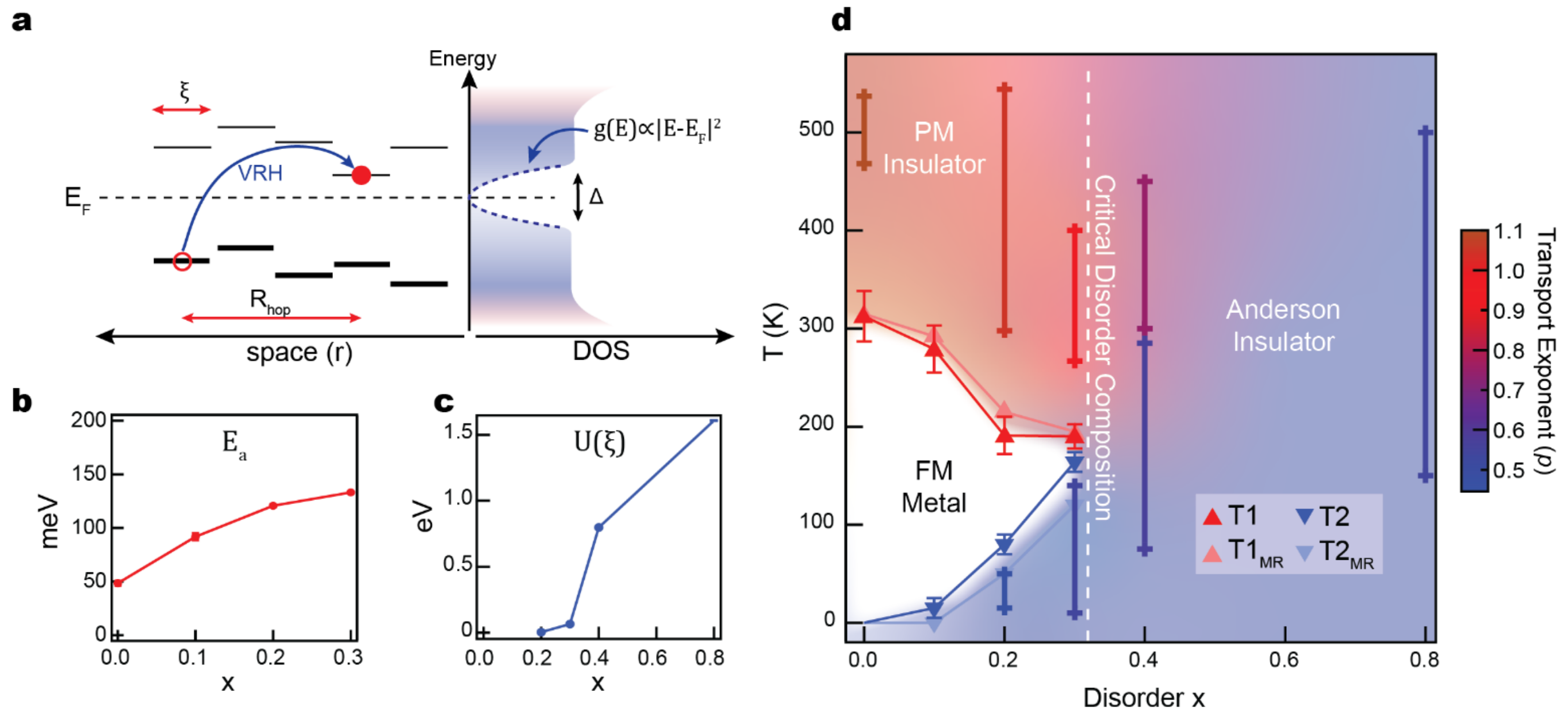

**Figure 4**. **(a)** Schematic for an interacting Anderson insulator showing the spatial and DOS distributions of localized states around an emerging soft Coulomb gap. **(b)** Extracted small polaron hopping activation energy ($E_a$) for the SPH and **(c)** Coulomb Energy evaluated at the localization length $U(\xi)$ for the VRH regions. **(d)** Proposed phase diagram including extracted transition temperatures (red and blue triangles) and generalized transport exponents. Marker color indicates the fitted transport exponent using the scale at right; vertical bars indicate the fitted temperature interval.

Next, we discuss a physical model of electronic transport in LS5B. As depicted in Figure 4a, the disordered crystal potential induces strong localization of energy states near the Fermi level. Due to the weakening of dielectric screening and strong electron correlations, double site occupancy is energetically unfavorable. In this case, transport is governed by the probability of finding a nearby empty site with a suitable energy. For ES-VRH, long range Coulomb interactions lead to a depletion of the DOS around the Fermi level, which gives rise to the soft Coulomb gap depicted in Fig. 4a[28,29]. This soft gap energy $\Delta$ is determined by the characteristic temperature $T_{ES}$ of the ES-VRH transport. Values of $T_{ES} \approx 25{,}000 - 52{,}000$ K are extracted directly from transport slopes of the highest substitution cases ($x$ =0.4, 0.8). See Supplementary S3 for details. We describe these gaps as "colossal", as such temperatures are nearly an order of magnitude higher (7.5×) than the highest previously reported high $T_{ES}$ values (up to ~7,000 K) in systems where granular transport can be clearly ruled out due to verification of a clean nanostructure[36,37]. A nanostructure free of grains is critical as Sheng-Abeles granular metal transport shares the same functional form as ES-VRH and leads to artificially inflated values of $T_{ES}$ due to the grain charging energy. Anomalously high $T_{ES}$ values (up to ~25,000 K) have been reported for some correlated oxides[38,39], however pure ES-VRH transport cannot be verified without microscopic analysis. For LS5B, the absence of a resolved structural granular network from XRD and HAADF-STEM indicates grain-charging does not influence the fitted $T_{ES}$ values. We refrain from comparing specific Δ values from transport as differing physical interpretations are used to define the gap and because the most rigorous evaluation requires measurement of $T_{ON}$ at the crossover to Mott-VRH with $p = 1/4$. We therefore report the screened Coulomb interaction energy, which is defined as $U(\xi_{loc})$, where the transport derived localization length is $\xi_{loc} = \frac{2.8\, e^2}{4\pi\epsilon_0\epsilon_\infty k_B T_{ES}} \approx 4.0 - 2.0$ Å for ($x$ =0.4, 0.8). Here, the dielectric screening is approximated from parent perovskites as $\epsilon_\infty \approx 5.0$ since hopping distance is confined within the volume of the film. Physically, this represents the largest electrostatic energy penalty of the system for hopping across the shortest distance, or $\xi_{loc}$. Calculations yield values as high as 1.6 eV for $x$ =0.8. The short $\xi_{loc}$ and $R_{hop}$ imply unit cell scale localization, consistent with a strongly localized Anderson insulator. In this regime, wavefunction overlap corrections extend the validity of ES-VRH[40,41].

From these results we construct a phase diagram of the LS5B system. The characteristic energies from transport equations are plotted in Figure 4b and 4c, with the corresponding small polaron activation energy $E_a$ 3for PMI and the $U(\xi_{loc})$ for I2. Employing the transition temperatures T1 and T2 obtained from resistivity and magnetoresistance, we plot the phase diagram in Figure 4d. Here, the transport exponent is given a color from p=0.45 (blue) to p=1.1 (red) and plotted by bars spanning the linear fit temperature range. From the transport (dark color triangles) and magnetoresistance (light color triangles) in Fig. 2e, we created the FMM-PMI and FMM-I2 boundaries for $x = 0.00 - 0.30$, while the transport exponents were used to determine the regions occupied by the I2 phase. A white dashed line approximates the Anderson disorder boundary. For the metallic regions $x < 0.30$, the system may exhibit a localized Anderson ground state, but thermal activation excites carriers across the Anderson mobility edge to an FMM state before the transition to PMI. We note that our low temperature measurements at $x < 0.30$ cannot completely rule out nanoscale electronic or magnetic phase separation leading to percolative transport. However, electronic transport hysteresis associated with domain nucleation and growth was not observed. Further, the HAADF-STEM images do not reveal structural domains, grain boundaries, or dislocations. A critical point emerges near $x = 0.30$, where T1 and T2 merge and the metallic state vanishes for all temperatures. We identify the compositions $x \geq 0.40$ as possessing an interacting Anderson state at all temperatures studied. We note that minor Sr and Cr enrichment of the surface (Supplementary S7) may shift transition temperatures relative to those expected from the nominal composition. Such surface effects are well documented in previous high-quality LSMO and HEO films[42–44]. We also note that short-range ordering (SRO) cannot be resolved by microscopy measurements and could have a small influence on the calculated disorder spacing $\xi_{dis}$.

**DISCUSSION**

Our results indicate LS5B is distinct from a superposition of its individual end member phases and therefore conduction is not governed by an element specific localization phenomenon but by a landscape of site-potential disorder. The single TM compounds $La_{0.7}Sr_{0.3}BO_3$ where B = (Mn, Co, Ni) show either bad metal characteristics or itinerant conduction with low room-temperature resistivity. In contrast, LS5B has an electrical resistivity at least one order of magnitude higher than the average of these compounds indicating that a simple rule of mixtures does not apply. Recent element specific measurements on related perovskites reveal that B-site substitution actively modifies local crystal fields, exchange pathways and spin states [32]. This composition-dependent redistribution of the local electronic environment provides a microscopic mechanism for the site-potential variance supporting Anderson localization. On the other hand, single B-site substitutions such as Cr, Co or Ni introduce competing effects such as short-range antiferromagnetic correlations, spin/valence disorder, or charge ordering[45–51]. Although insulating behavior emerges in some cases, these substitutions do not establish Anderson behavior over broad temperature ranges. For the case of $La_{1-x}Sr_xMn_{1-y}Fe_yO_3$, the substitution of Fe suppresses Mn-based double exchange and drives the system from a ferromagnetic metal toward a localized conductor with a small region of transport that includes ES-VRH with $p = 0.5$, along with many other localization mechanisms and exponents[52,53]. However, due to the polycrystalline, granular nature of the samples and the wide variety of transport mechanisms observed in previous studies, it is difficult to make direct comparisons to LS5B. In summary, these single element substitutions highlight a broad range of localization effects but are qualitatively distinct from LS5B and generally do not present interacting Anderson insulator behavior.

We present multiple lines of evidence that establish Anderson localization is intrinsic to the transition metal cation sublattice of LS5B. This is critical, because another source of Anderson behavior is that of charge compensation induced by extrinsic oxygen vacancy defects which create atomic-scale site disorder when reaching large concentrations. For example, it was observed previously that extremely oxygen deficient HEO with rare-earth element complexity can exhibit Anderson effects[54] but this is also commonly observed for other defective transition metal oxide systems[17,55]. Oxygen-rich growth and post-growth annealing, together with the absence of a systematic change in the Mn valence from XPS, argue against a simple Mn-reduction driven explanation for LS5B. These measurements do not provide a complete oxygen or multielement charge balance, and minor contributions from oxygen non-stoichiometry cannot be entirely excluded by them. However, our ARPES Fermi surface analysis demonstrates that the hole pocket carrier density remains constant at ~0.45 holes/unit cell for $x = 0.00 - 0.20$ (see Supplementary Text S6). Full compensation of the nominal Sr-induced holes by oxygen vacancies would require $\delta = 0.15$, reducing the nominal B-site valence from 3.3+ to 3+. Beyond $x = 0.00 - 0.20$, a large deficiency would lead to a reduction in average valence state of the TMs which should be captured by XPS. However, XPS confirms stable Mn valency across the LS5B series (See Supplementary S5). Further, our experiments indicate a systematic change in localization length with LS5B composition rather than being fixed by another type of defect density such as low angle grain boundaries or dislocations. For example, a calculation of $\xi_{loc} \approx 4.0 - 2.0$ Å for $x$ =0.4-0.8 agrees qualitatively with disorder length $\xi_{dis}$ as inferred from composition (Fig. 1b). In summary, we propose Anderson behavior is native to LS5B and defects or grains are not responsible.

Next, we discuss when to anticipate Anderson behavior more broadly in HEO. The LS5B system possesses highly dispersive 3D perovskite bands and therefore has the least sensitivity to site potential variance $\Delta V_i$ due to its large bandwidth. Therefore, Anderson localization should be widely observed in other HEO with transition metal complexity since most structures have lower bandwidth, such as spinels. For perovskites there is large orbital overlap yielding approximate bare bandwidths $W = 1.0 - 1.2$ eV with transition metal potential variance at $\Delta V_i \approx 1 - 2$ eV. In this case, substantial substitutions are needed $x \geq 0.30, 0.40$ to meet Anderson criteria at room temperature ($V_i/W > 1.5$)[56,57] as we reported with LS5B. However, for spinel structures the bandwidth $W$ is reduced to approximately 0.3-0.5 eV and therefore Anderson behavior should be seen at lower disorder levels[58,59]. Extreme examples can be found in layered systems, where localization effects are strengthened by the lower dimensionality. The layered oxide $LiCoO_2$ has significantly reduced band structure dimensionality and has already been proposed to undergo Anderson localization at low defect concentrations[60]. Another example is the layered cuprate $LiCu_3O_3$, where defect-induced ES-VRH and suppressed spectral weight at the Fermi Energy is observed[36]. Therefore, we anticipate HEO layered structures to exhibit Anderson behavior at the lowest substitution levels. A future line of work is to create a theoretical framework wherein Anderson behavior is predicted to emerge more broadly in HEO or CCM from basic criteria such as bandwidth, site potential variance and substitution level.

Native Anderson localization has technological implications for HEOs across many applications[12–15,61]. In photocatalytic HEO particles, photo-excited carrier migration to the surface will be bottlenecked by the Anderson localization length rather than the band gap; optimizing catalytic rates will therefore require tailoring particle sizes to the stoichiometry-dependent disorder length. Similarly, complex layered oxide battery cathodes (e.g., Ni-Mn-Co) offer high theoretical capacity but often suffer from poor rate capability. As chemical de-intercalation shifts the Fermi level into disordered bands, Anderson localization can trap carriers before they reach conductive fillers. Rather than relying on traditional doping, battery designers may need to tailor particle dimensions to mitigate localization. Conversely, Anderson effects can be advantageous in high-field applications like ferroelectric capacitors. While HEO capacitors demonstrate extraordinarily high energy density and breakdown resilience, the mechanism remains unclear[62]. We propose that under high operating fields (>1 MV/cm), HEO Anderson insulators may prevent carriers from gaining sufficient kinetic energy for Poole-Frenkel or Fowler-Nordheim assisted avalanche breakdown. Engineering shorter Anderson localization lengths thus provides a novel modality to suppress extended conduction and maximize recoverable energy.

## CONCLUSION

LS5B hosts Anderson behavior in ambient conditions without gross defects or grain boundaries meaning quantum mechanical interference of electron wave packets is a default characteristic. This stands in direct contrast to nearly all other material classes in which only specific defect concentrations, substitutions or nanostructure result in quantum mechanical interference. Our results indicate that HEO naturally exhibit high site potential variations $V_i$ that exceed the bandwidth $W$ and meet the Anderson criteria for destructive quantum interference ($V_i/W > 1.5$). Given that transition metal complexity is a common design strategy, we anticipate HEO to exhibit Anderson behavior generally. Our phase diagram for LS5B provides the beginning of a framework for HEO design, with implications for the broader class of CCM.

**METHODS**

Duty cycling between targets of $La_{0.7}Sr_{0.3}MnO_3$ and $La_{0.7}Sr_{0.3}(Mn, Cr, Fe, Co, Ni)_{0.2}O_3$ was used for the gradual composition changes in LS5B. Sub monolayer shots from each target was cycled to achieve high intermixing. X-ray diffraction and reflectivity measurements were performed after growth to check phase and thickness. All samples were grown using a 248 nm laser and a fluence of 2.7 $J/cm^2$. Samples were annealed in air at 500 °C after growth to reduce oxygen defects. For photoemission studies, a thin underlayer of metallic LSMO is grown under the 10 nm film to reduce energy shifts due to photoemission induced charging. Electronic transport, magnetoresistance and magnetization curves are done in a Quantum Design Physical Property Measurement System (PPMS). Electronic measurements used van der Pauw configuration. For measurements between 5 K and 400 K, the samples were kept in a clean He atmosphere around 0.8-10 Torr. For high temperature measurements up to 570 K the transport was performed in air to avoid oxygen vacancy formation. Magnetization curves were performed on thicker 50 nm films to obtain enough volumetric signal. Angle Resolved Photoemission was performed at Maestro and Merlin Beamlines at ALS. Photon energies used are in the 100-150 eV range, with the reported data in this main text corresponding to 131 eV. The scattering plane was aligned to one of the in-plane axes of the film. Data shown in the main text were taken with linear horizontal polarization. STEM-EDS measurements were performed on Ga-FIB prepared TEM lamellae (Thermofisher Helios 660 Nanolab) on a ThermoFisher Titan Themis Z operated at 300 kV. Beam currents between 150 and 350 pA were used with pixel dwell times ranging from 10 µsec to 10 msec. For shorter pixel dwells, multiple passes of each pixel were summed to get appropriate signal to noise. Beam damage and acquisition artifacts were observed. For the final analysis, single-pass acquisitions were used with pixel dwell times of approximately 5 msec or multipass acquisitions with pixel dwell times of 10 μsec. For multipass datasets, frames showing specimen damage (e.g., holes) or image/acquisition artifacts (e.g., smearing or scan jumps) were excluded before summation.

## ACKNOWLEDGMENTS

This work was supported as part of the Center for Reconfigurable Electronic Materials Inspired by Nonlinear Neuron Dynamics (ReMIND), an Energy Frontier Research Center funded by the US Department of Energy (DOE), Office of Science, Basic Energy Sciences. E.S. and E.J.F. were additionally supported by the Sandia Laboratory-Directed Research and Development (LDRD) Program. Sandia National Laboratories is a multimission laboratory managed and operated by National Technology and Engineering Solutions of Sandia, LLC., a wholly owned subsidiary of Honeywell International, Inc., for the US Department of Energy's National Nuclear Security Administration under contract DE-NA-0003525. H.M.H. and Y.L. were supported by a Sandia University Partnership Network contract under the Sandia LDRD program. This research used resources of the Advanced Light Source, which is a DOE Office of Science User Facility under contract no. DE-AC02-05CH11231. This paper describes objective technical results and analyses. The views expressed in the article do not necessarily represent the views of the US Department of Energy or the United States Government. We thank Enrique Garcia Michel, Arantzazu Mascaraque and Miguel-Ángel González Barrio for their critical reading of the manuscript and insightful discussions regarding magnetic measurements.

**Supplementary Text For:**

# High Entropy Oxides as Room Temperature Anderson Insulators with Colossal Soft Gaps

*Elena Salagre[1], Heather Hare[2], Eli Rotenberg[3], Jonathan D. Denlinger [3], Joshua D. Sugar[1], Yiyang Li[2], and Elliot J. Fuller[1]**

[1]Sandia National Laboratories, 7011 East Ave. Livermore, CA 94550, United States of America
[2] Materials Science and Engineering, University of Michigan, Ann Arbor, MI, 48109, United States of America
[3]Advanced Light Source, Lawrence Berkeley National Laboratory, Berkeley, California 94720, United States of America

## Table of Contents

## S1. Single Element Ground States

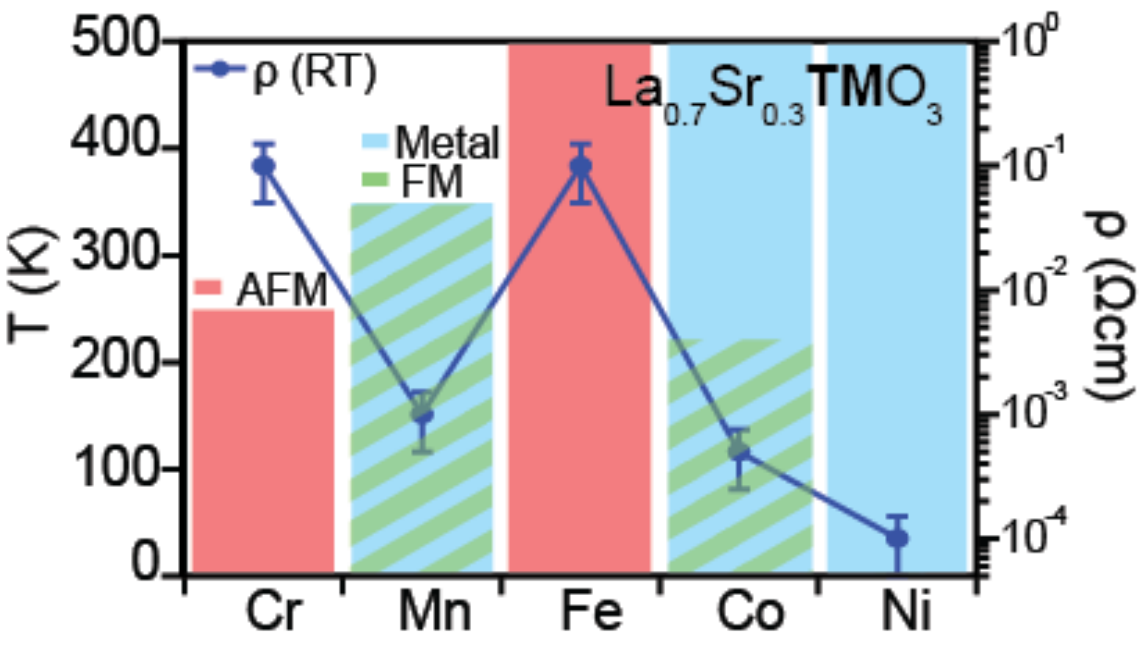

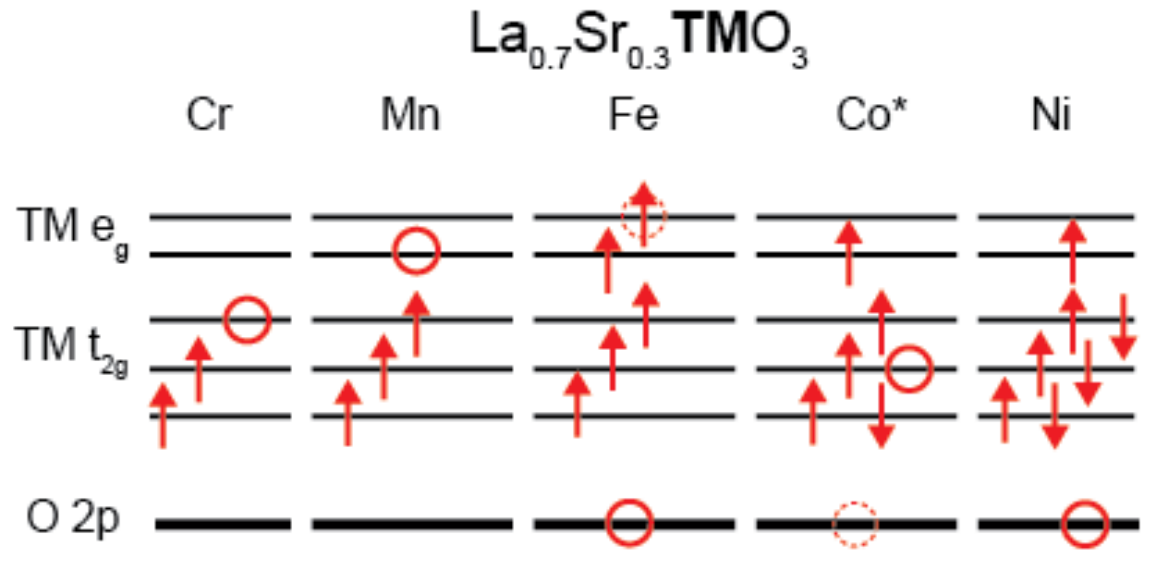

**Figure S1.1** Schematics of parent compound behavior and orbital occupation.

In Figure S1.1 the general behavior of $La_{0.7}Sr_{0.3}TMO_3$ and general orbital occupation are shown. The uncertainty bars in the resistivity values come from differences in sample morphology and quality and can also shift transition temperatures. The Cr and Fe compounds show AFM at low temperature, with relatively high resistivity values. On the other hand, Mn and Co compounds show FM character and large temperature ranges occupied by a metallic phase. The case of the Ni base system is an outlier, as it shows no magnetic response and enduring metallic behavior. The series LS5B exhibits a gradual change in band occupation, band gap and charge transfer energy. In addition to the discussed disorder induced localization and the FM network disruption, other effects can influence the local environment of the system. For example, contributions from short range order and substrate strain can introduce variations in the orbital occupancy and preferred valence state of each of the TMs. We note substrate strain variations were minimized to less than 1% during growth.

## S2. X-ray diffraction

Additional XRD for dilute substitutions is shown in Figure S2.1a with extracted lattice parameters and out of plane and in plane strain compared to STO (Figure S2.1b). Note that the in-plane strain has been calculated using a constant volume approximation. The films used for structural determinations are 50 nm thick and therefore are mostly relaxed. Examples of XRD of the actual 10nm films discussed in this study are shown in Figure S2.1c.

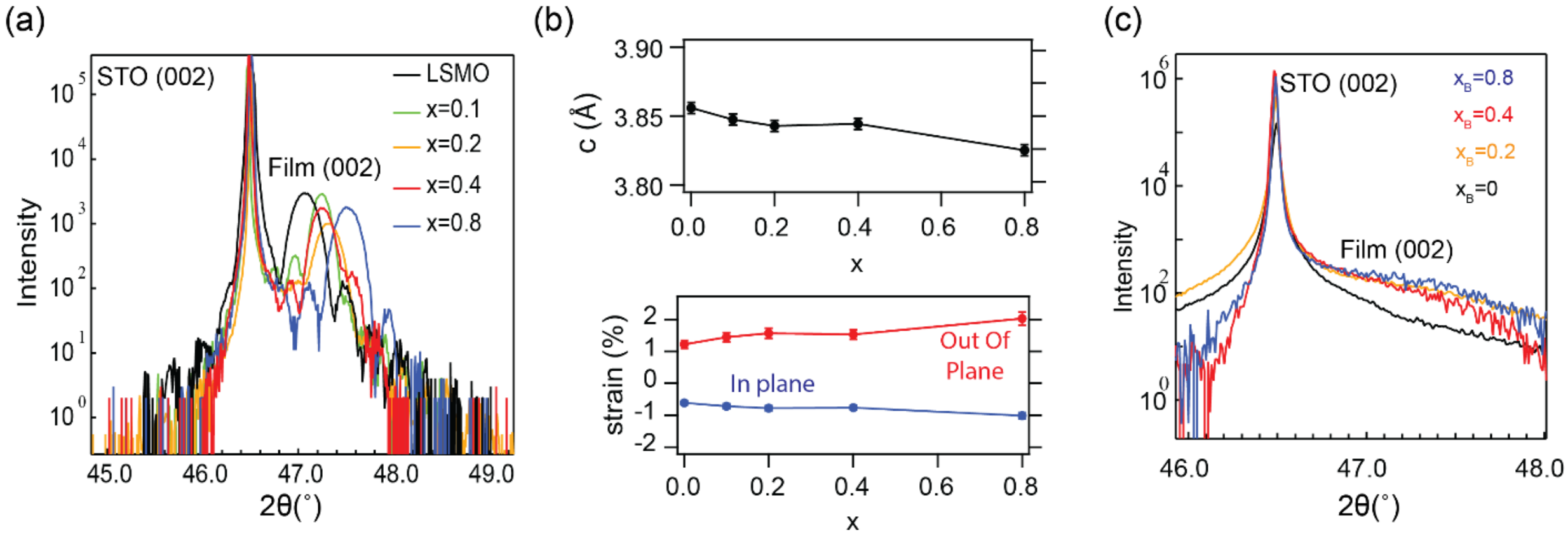

**Figure S2.1** (a) XRD of the (002) pseudocubic peak for 50 nm films on STO and (b) extracted lattice and strain values. (c) XRD of the 10 nm films.

## S3. Electronic Transport Fitting

The electronic transport was studied using the generalized transport discussed in the main text. Variable range hopping between states around the Fermi level is expected, with competition between tunneling and activation. The most unbiased analysis is done by the Zabrodskii to obtain the exponent that indicates the dominant transport mechanism. Given the strong Coulomb interactions expected in our system and the obtained exponent of ½, the additional electrostatic interactions must be introduced, yielding the ES-VRH dominant mechanism. These interactions induced a progressive reduction of populated states around EF, following the relationship $N(E) \propto E^2$ for a 3D system. The transport can be used to extract the characteristic temperature $T_{ES}$ and estimate the energy scale of the soft gap. Plotting $\ln(\rho)\ vs\ T^{-1/2}$ and performing a linear fit, we obtain $T_{ES} = m^2$ , with m being the slope of the linear fit. To estimate the characteristic hopping-energy scale, we consider a hopping exponent containing tunneling and activation terms, $\frac{2R}{\xi} + \frac{E(R)}{k_B T}$, with R hopping distance and $E(R) \propto \frac{C}{R}$ with C being a constant related to the dielectric screening and dimensionality. Minimizing this energy and expressing the characteristic temperature as $T_{ES} \propto \frac{C}{k_B \xi}$ we obtain

$$\varepsilon_{hop}(T) \propto \frac{1}{2} k_B \sqrt{T_{ES} \cdot T}$$

An evaluation at the crossover temperature between Mott-VRH and ES-VRH is usually used to determine the soft gap, but it is not valid for the ES-VRH-SPH or ES-VRH - metal transition cases. This is because it relies on the thermal fluctuations that overcome the soft gap leading to tunneling with p=1/4 Mott-VRH. However, given the large TES obtained and the Coulomb energy discussed in the main text.

Extended temperature range transport is shown in Figure S3.1 a, this data is also used for the high temperature exponent change discussed in the main text and the Zabrodskii plot shown in Figure S3.1c. The Metallic region has been fitted with a double exchange transport model:

$$\rho(T) = \rho_0 + A \cdot T^2 + B \cdot T^{9/2}$$

Where the $T^2$ accounts for electron-electron scattering and $T^{9/2}$ is characteristic of two-magnon scattering processes typical in double exchange manganites.

The results are shown in Figure S3.1b and summarized in Table S3.1

| | $\rho_0$ (Ω·cm) | $A$ (Ω·cm·K$^{-2}$) | $B$(Ω·cm·K$^{-9/2}$) |
|---|---|---|---|
| **x =0.00** | $(1.63\pm0.13)\cdot10^{-3}$ | $(4.37\pm0.72)\cdot10^{-9}$ | $(1.07\pm0.52)\cdot10^{-15}$ |
| **x =0.10** | $(6.71\pm0.22)\cdot10^{-3}$ | $(1.09\pm0.16)\cdot10^{-8}$ | $(3.60\pm0.16)\cdot10^{-14}$ |
| **x =0.20** | $(6.18\pm0.05)\cdot10^{-1}$ | $(-9.27\pm0.54)\cdot10^{-7}$ | $(4.14\pm0.10)\cdot10^{-12}$ |

**Table S3.1**. Results of Fitting to double exchange transport model.

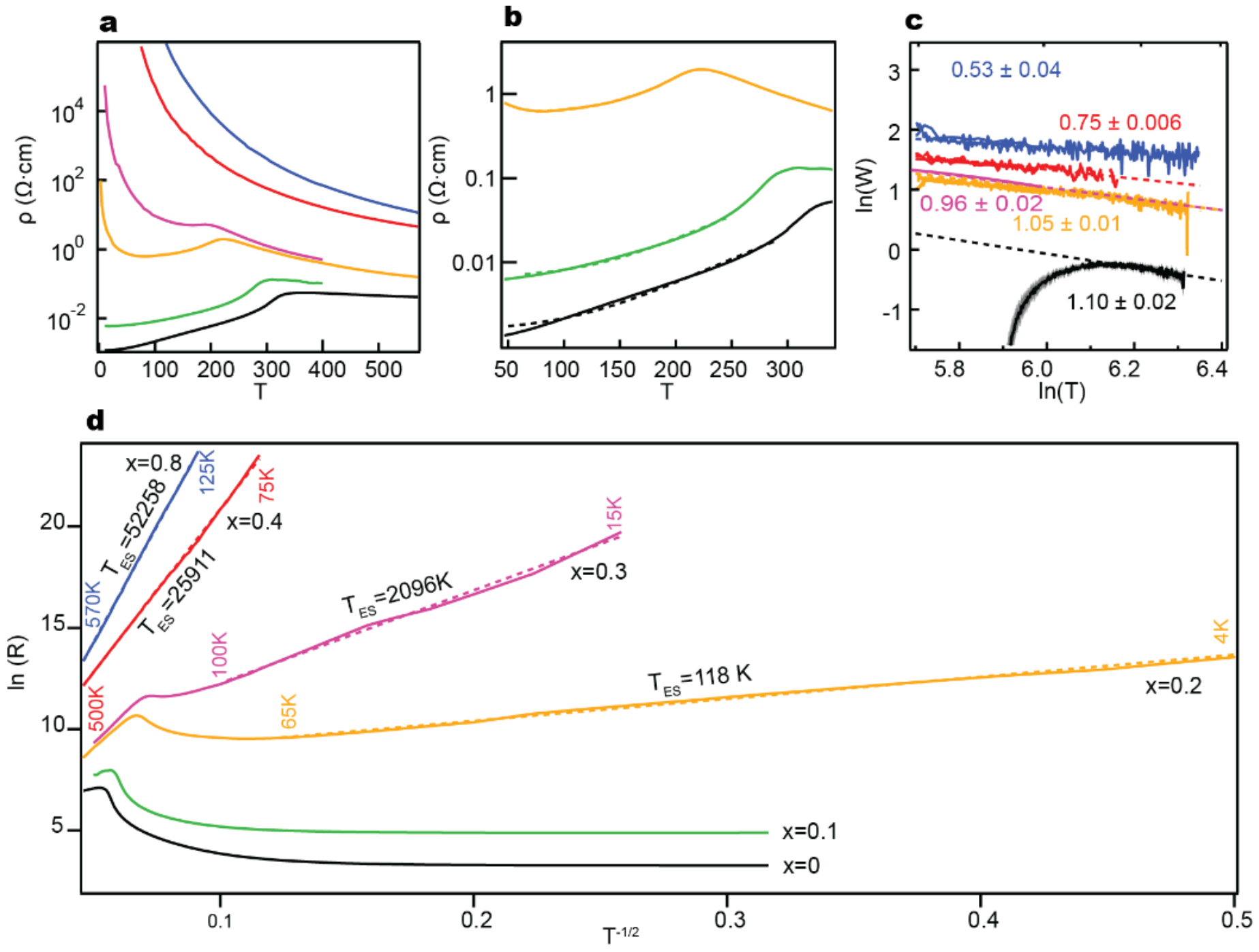

**Figure S3.1** Supplementary electronic transport data. (a) Extended high temperature range for selected samples. (b) zoom and fits of the metallic region. (c) Zoom and fit of the Zabrodskii plot high temperature region showing SPH and high temperature ES-VRH regimes, parent compounds. (d) ln(R) vs $T^{-1/2}$ linear fits to extract $T_{ES}$ and related metrics.

The metallic regions observed, fitted to a double exchange transport model as shown in Fig. S3.1b, suggest that the initial LSMO metal, consistent with previously reported films, increasingly diverges from this model as $x$ increases, with signatures of stronger residual scattering disrupting the double exchange mechanism[63,64].

The transition temperatures and widths for T1 have been obtained using a sigmoidal fit of the transition regions, of the form $\rho(T) = \rho_{min} + \frac{\rho_{max}}{1+\exp(\frac{T_1-T}{r})}$ and the 10-90% transition width is taken as $\Delta T = 2 \cdot \ln(9) \cdot r$. Given the low range and steepness of the low temperature T2, the change in slope from resistivity is used to determine the transition temperature and the width is estimated from the error.

A ln(R) vs $T^{-1/2}$ is shown in Figure S3.1d and used to obtain the linear fit ranges and characteristic $T_{ES}$ temperature. Extracted characteristic temperatures $T_{ES}$, and Coulomb localization energies $U(\xi) = \frac{k_B T_{ES}}{2.8} = \frac{e^2}{4\pi\varepsilon_0\varepsilon_\infty\xi}$ are summarized in Table S3.2. Other transport models have also been fitted to the data to compare fitting quality and range, this is shown in Figure S3.2 for Arrhenius activation and adiabatic polaron hopping, we note that, comparing to Fig. S3.1, the emergent I2 phase is better interpreted as ES-VRH for the low temperature range, as expected from the reduced transport exponent analysis. Fitting ranges are indicated for each case. From polaron hopping fits, the activation energies shown in Fig. 4 were extracted and are included in Table S3.2. Additionally, the localization length $\xi_{loc} = \frac{2.8\, e^2}{4\pi\epsilon_0\epsilon_\infty k_B T_{ES}}$ and $R_{hop}(T) = \frac{1}{2}\,\xi_{loc}\sqrt{\frac{T_{ES}}{T}}$ at the highest temperature of the fitted range are also shown in Table S3.2. We note that in all cases except x=0.2 the hopping distance is in the order or shorter than the thickness of the film (~100 Å). For the specific case of x=0.2, the VRH physical validity may not be unique and other contributions to transport may be limiting the validity of this interpretation. The classic criteria for ES-VRH validity, $\frac{R_{hop}}{\xi_{loc}} \leq 1$ , is not an invalidating criterion when T is close to $T_{ES}$ or when the system is close to a critical regime.  We also note that no error due to the uncertainty in $\varepsilon_{\infty}$.

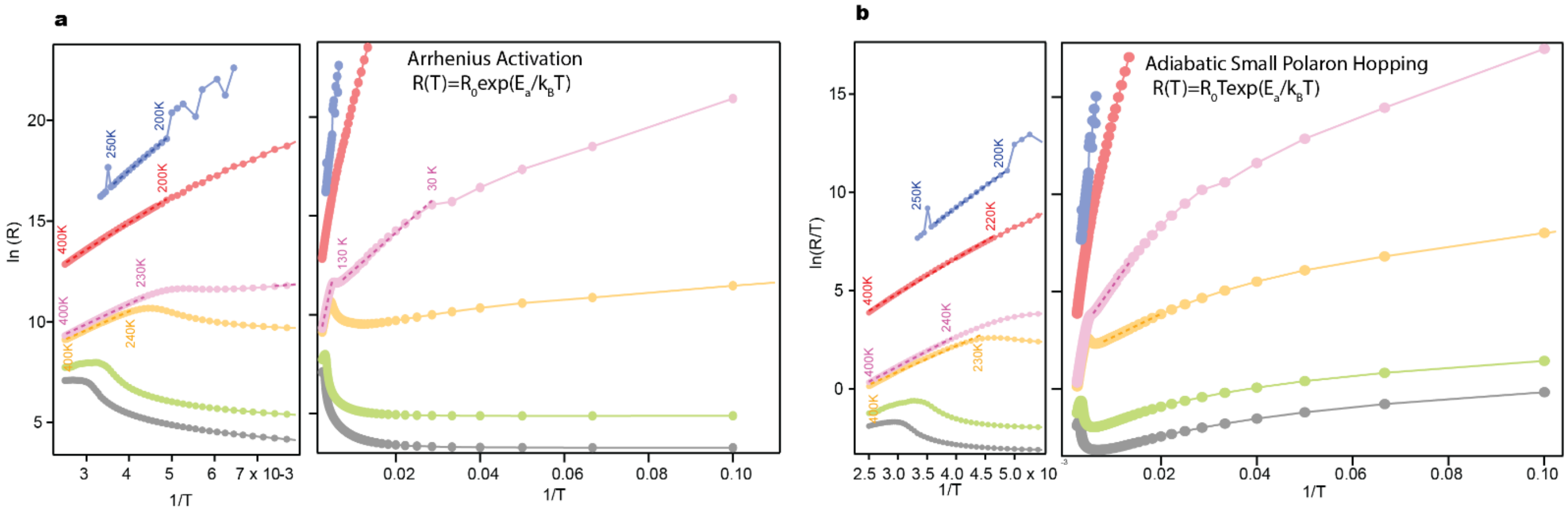

**Figure S3.2** Additional model fits for the transport data.

| | $T_{ES}$ (K) | $U(\xi)=k_B T_{ES}/2.8$ (meV) | $E_a$ (meV) | $\xi_{loc}$ (Å) | $R_{hop}(T)$ (Å) |
|---|---|---|---|---|---|
| **x=0.0** | - | - | 48.50±2.21 | | |
| **x=0.1** | - | - | 91.84±3.26 | | |
| **x=0.2** | 118±3 | 3.63±0.11 | 120.71±1.28 | 792.99±20.20 | 534.23±6.80 |
| **x=0.3** | 2096±68 | 64.50±2.09 | 133.13±0.79 | 44.64±1.45 | 102.19±1.70 |
| **x=0.4** | 25911±42 | 797.41±1.29 | - | 3.61±0.01 | 13.00±0.01 |
| **x=0.8** | 52258±73 | 1608.23±2.25 | - | 1.79±0.03 | 8.57±0.01 |

**Table S.3.2** Summary of electronic transport extracted metrics.

## S4. Magnetic Measurements

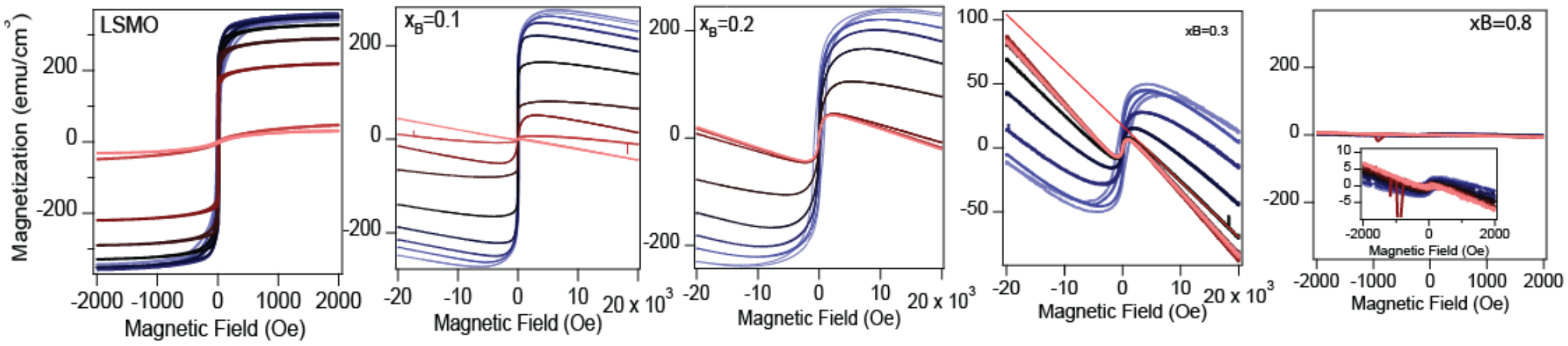

**Figure S4.1** Extended range without substrate contribution subtraction of the magnetization data.

Because the magnetic moments of the 9–10 nm films are near the sensitivity limit, magnetization loops were measured on companion 50 nm films grown under identical conditions. These measurements are used to track the composition-dependent collapse of ferromagnetism, while the phase boundaries are assigned primarily from transport and magnetoresistance measured on the thin films. Magnetization loops are plotted here without background subtraction. Background slopes are expected due to the well-known contributions to magnetization loops from the substrate.

## S5. X-ray photoelectron spectroscopy

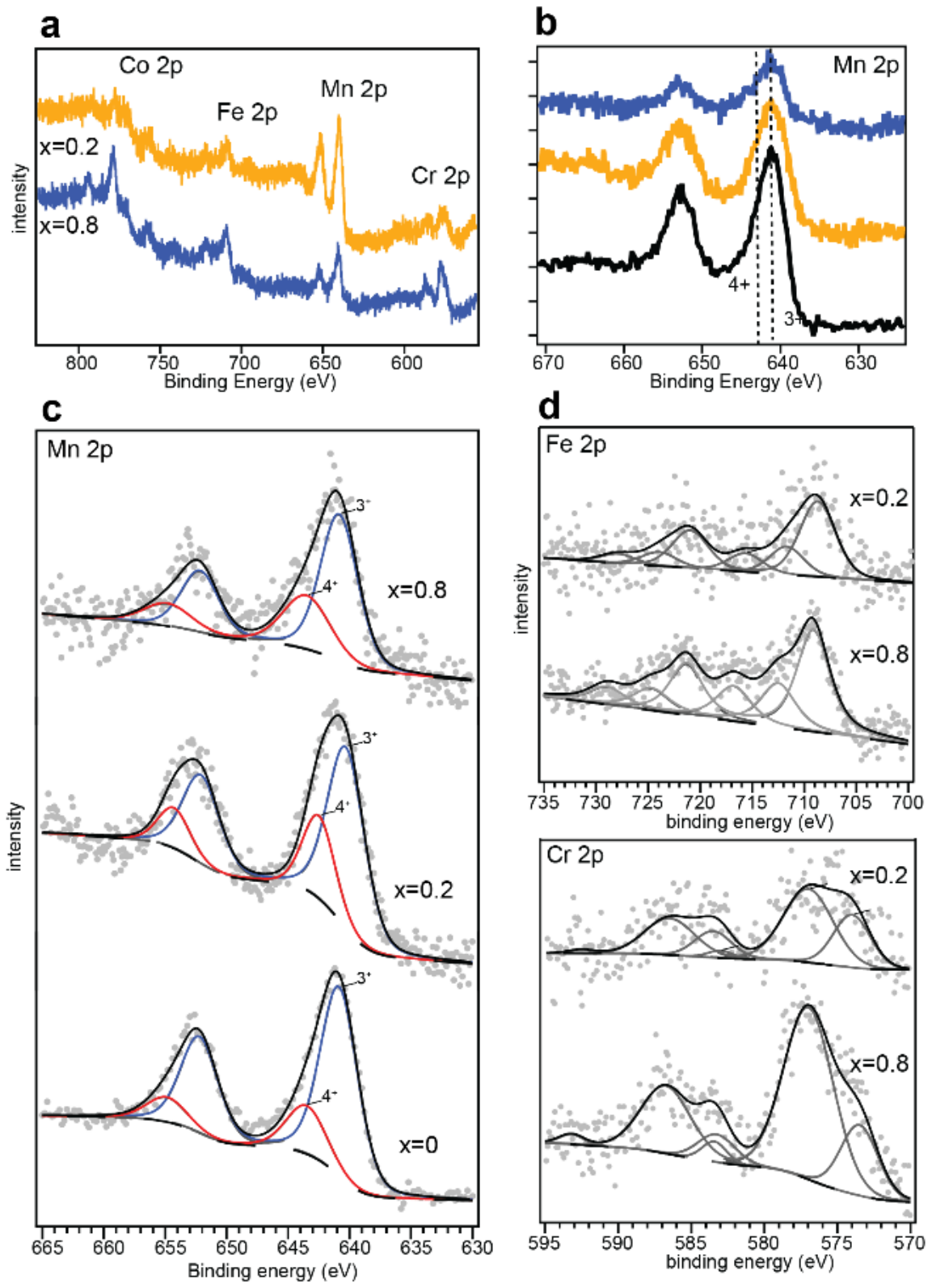

**Figure S5.1** (a) Transition metal 2p core levels for x=0.2 and x=0.8. (b) Mn 2p for x=0, 0.2 and 0.8 showing the evolution of core level position and asymmetry. (c) Deconvolutions for Mn 2p shown in (b). (c) Deconvolution fits for Fe 2p and Cr 2p. Note that Co and Cr overlap with Auger features. Mn shows a similar Mn4+/3+ ratio for all compositions within error.

XPS of the x=0, 0.2 and 0.8 samples is shown in Figure S5. The fits of Mn 2p $Mn^{3+}$ and $Mn^{4+}$ don't show significant ratio changes in the multi-valence ratio, suggesting that Mn role in the lattice does not evolve and that there is no significant increase in oxygen vacancies accommodated by the Mn. Transition metal area ratios are consistent with the expected stoichiometry for dilute and heavy substituted samples. Fe 2p and Cr 2p have also been fitted to a multivalency deconvolution. The overlap of Ni 2p and Co 2p with La 3d and MNN Auger peaks prevents a good fit in those cases. Fe 2p shows a consistent valence ratio for a 30% Sr substituted compound. For the x=0.8 sample, Cr shows an increased high binding energy component typically associated with $Cr^{4+}$, which could be linked to Cr and Sr enrichment in the surface observed in EDS (see S7). We note that the Cr enrichment and multivalency for the x=0.2 case is not sufficient to explain the observed ARPES features and therefore electronic localization is still the dominant origin. These results have been obtained using Al Kα (1487 eV) and show no apparent sign of transition metal segregation at the surface layer within the technique resolution.

The probing depth of the ARPES measurements is estimated to be ~1 nm while that of XPS is approximately ~5 nm, however, photoelectron signal decreases exponentially with depth, with the surface contributing strongly to the measured intensity. The observed Sr enrichment in LSMO has been extensively reported and Cr surface diffusion or changes in Cr valency are reported in HEOs. Such changes do not undermine the disorder-localization interpretation as they preserve/enhance potential disorder, but they may affect quantitative spectral weights. Moreover, the observed $k_z$ dispersion of the $e_g$ band and the absence of surface state signatures, indicate that the band structure measured is bulk-like. Additionally, ARPES constant energy surfaces do not show signs of surface reconstruction that could locally change the system's symmetry.

## S6. Additional ARPES data and Analysis.

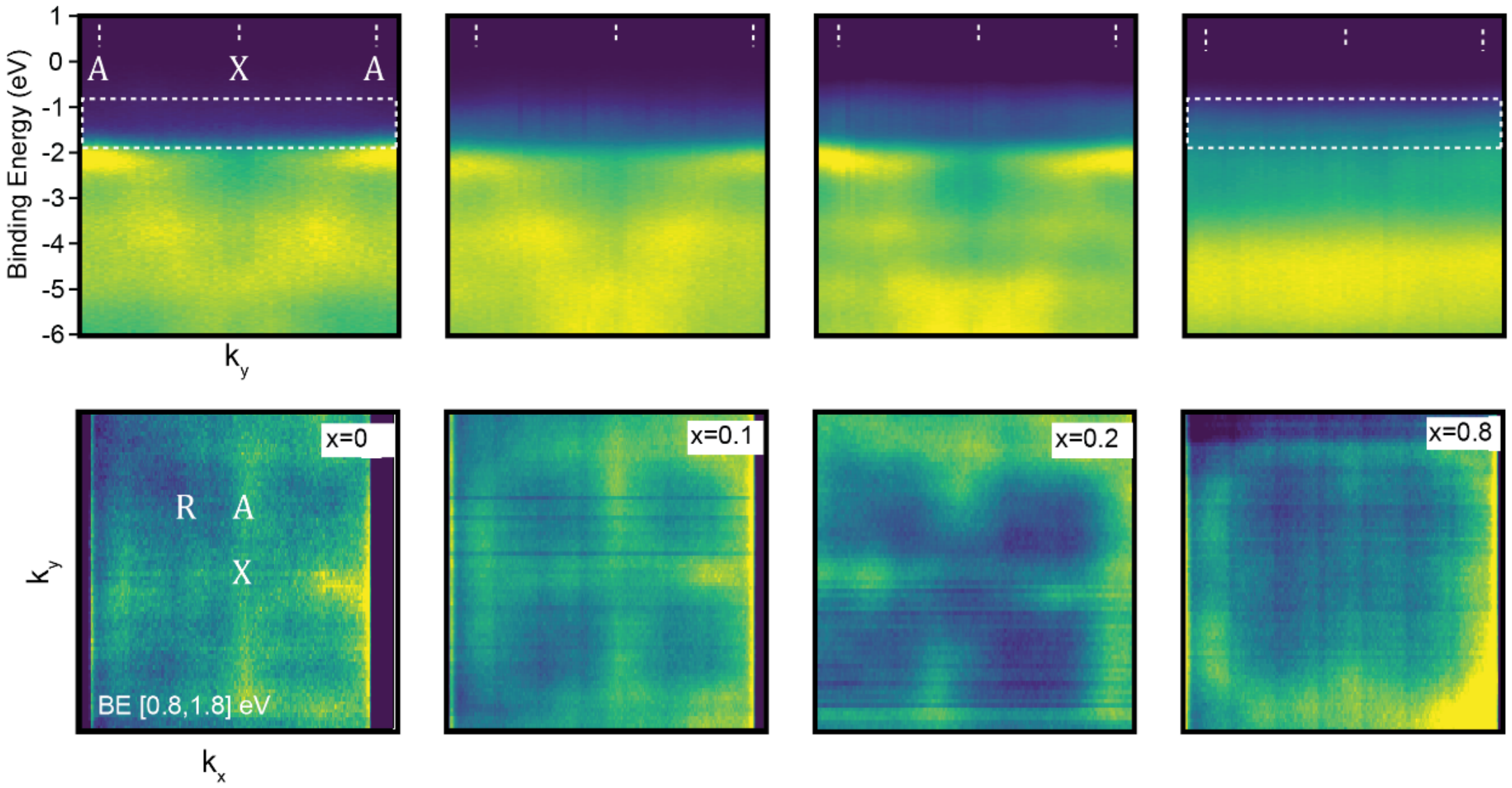

**Figure S6.1** Extended energy range for the valence band carpets of x=0, 0.1, 0.2 and 0.8. Despite the spectral weight changes, dispersion of the deeper bands is partially conserved, as shown in the bottom integrated constant energy surfaces.

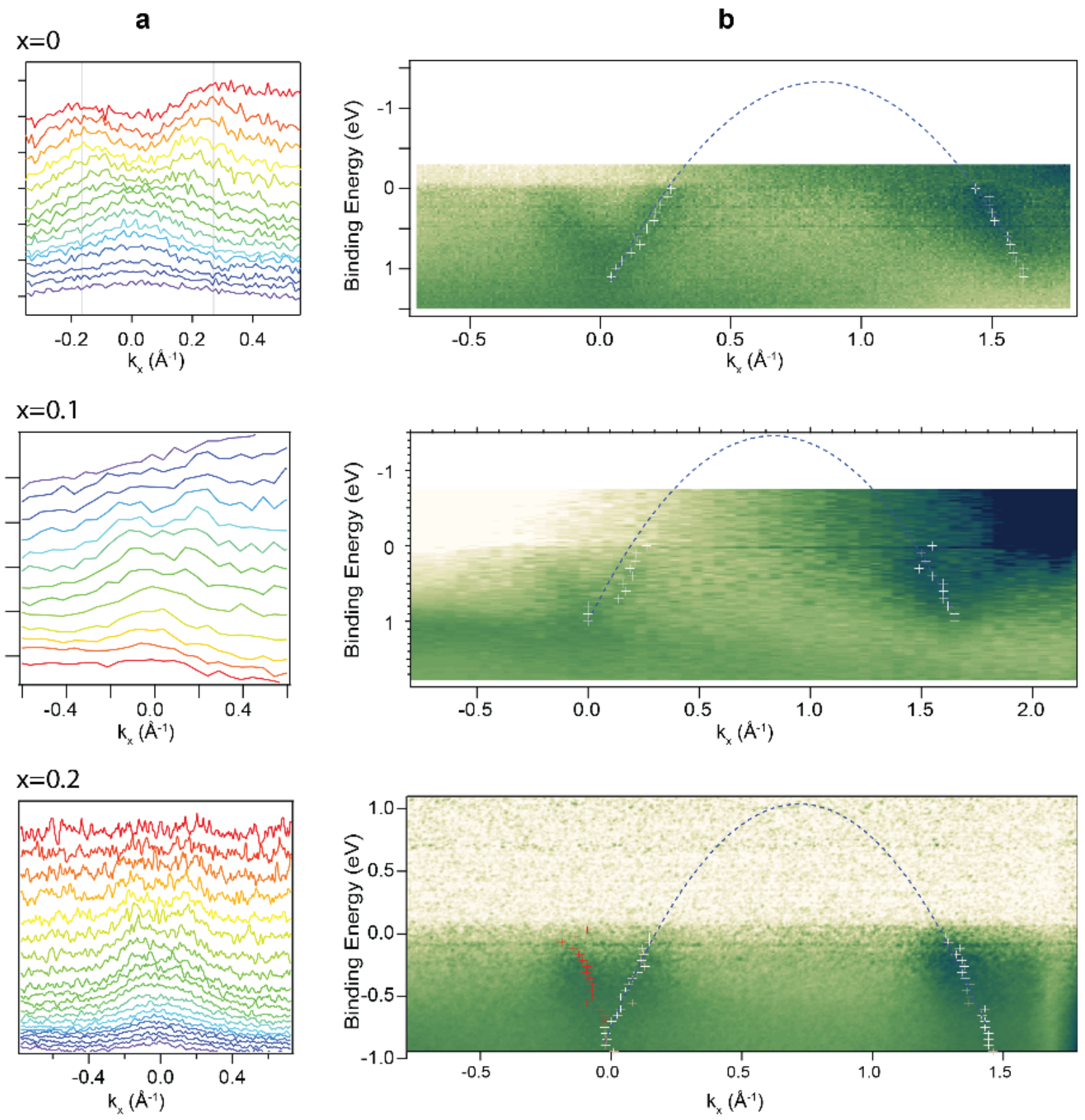

**Figure S6.2** Example of curvature analysis. Values in Table S6.1 are obtained by averaging MDC derived positions and FWHM of the observed crossings of the bands in along $k_x$ and $k_y$ R-A-R cuts.

Curvature analysis of the hole pocket feature has been performed taking into account the expected band shape of LSMO. Using a parabolic dispersion, the band can be fitted using

$$\frac{1}{m^*} = \frac{1}{\hbar^2}\frac{\partial^2 E}{\partial k^2}$$

The band was fitted using MDCs at different energies, as shown in Figure S6.2a and b. Similarly MDCs in $k_x$ and $k_y$ reciprocal space directions have been used to obtain an average value for the size of the hole pockets, examples are shown in  Figure S6.2c. Hole Density (hole/unit cell) and net carrier density has been extracted using Luttinger's theorem considering a cubic-like approximation of the hole pocket and spherical electron pocket, and therefore is expected to have some deviation from the nominal values, as despite clear 3D dispersion, the full shape of the hole pocket was not mapped. This hole density is accompanied by a small electron pocket, both spin-polarized. The electron pocket is not clearly visible at the main photon energy used but photon energy dependence and estimations from the diffuse intensity observed in the Fermi surfaces suggest an approximate radius of 0.2 Å$^{-1}$ giving a net carrier count shown in the third column of Table S6.1 where this value has been chosen constant within resolution error. This number gives a good estimation for the lack of change in carrier density during substitution.

$$p_h = \frac{V_{hp}}{V_{BZ}}\ ;\ V_{BZ} = \left(\frac{2\pi}{a}\right)^3\ ;\ V_{hp} = (2\cdot k_F)^3$$

The fits are shown in Fig S6.2 and obtained values for effective mass and hole density based on size of the hole pocket are summarized in Table S6.1 and shown in the main text in Fig.3.

| | $m_h*/m_0$ | $p_h$ | $p_{net}$ |
|---|---|---|---|
| **x=0.00** | 1.10± 0.10 | 0.41±0.08 | 0.36±0.10 |
| **x=0.10** | 1.08± 0.17 | 0.49±0.11 | 0.44±0.12 |
| **x=0.20** | 1.07 ± 0.15 | 0.42±0.13 | 0.37±0.14 |

**Table S6.1.** The effective mass and carrier density from fittings in Fig. S6.2

Observed bandwidth, as described in the main text, is defined as $W_{ARPES} = E_F - E_{min}^{eg}$ .

For the analysis of the electronic localization the Ioffe-Regel limit has been experimentally obtained. Following the criteria;

$$k_F l \sim \frac{k_F}{\Delta k} \rightarrow 1$$

Where $k_F$ is the Fermi wavevector of the hole pocket obtained from MDC in $k_x$ and $k_y$ direction along the R-A-R. $l$ is the mean free path of the quasiparticle and is estimated using the relationship with the momentum width of the band crossing at $E_F$ as $l = \frac{1}{\Delta k}$. In order to avoid artifacts due to data normalization the values in Table S6.2 were obtained from averaging the fits over the observed cuts in the raw Fermi surface cuts shown in Figure S6.3. The uncertainty has been obtained by error propagation using the fit uncertainty in the MDC fits. We note that experimental broadening and $k_z$ broadening have not been considered for error determination, since these would be constant factors in the x series, the trend is conserved, suggesting the approach to a localization limit. Owing to the finite photoelectron depth, the out-of-plane $k_z$ resolution is limited. Appreciable photon energy dependence of the $e_g$ states was observed probing the three-dimensional nature of the valence band. This dispersion may contribute to the spectral linewidth and explain the small electronic mean free path in LSMO.

| | $k_F$ [Å$^{-1}$] | $\Delta k$ (FWHM) [Å$^{-1}$] | $k_F l$ |
|---|---|---|---|
| **x=0.00** | 0.59±0.06 | 0.16±0.04 | 3.69±0.99 |
| **x=0.10** | 0.63±0.07 | 0.28±0.05 | 2.24±0.48 |
| **x=0.20** | 0.60±0.09 | 0.48±0.15 | 1.25±0.44 |

**Table S6.2.** Values for Fermi wavevector and FWHM obtained from MDC. Extracted $k_F l$ are also shown.

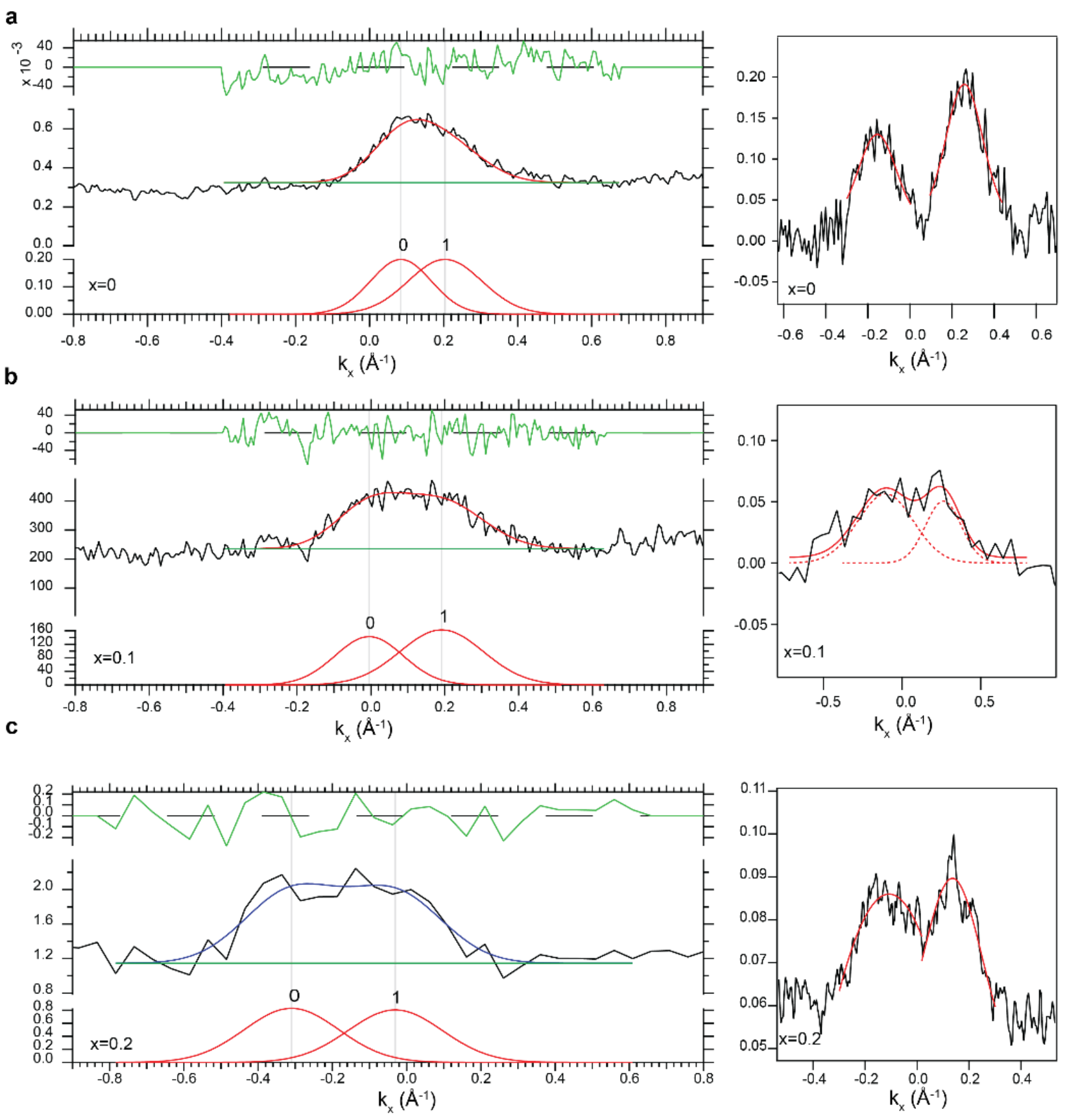

**Figure S6.3** MDC fits of kx and ky cuts along RAR direction to obtain Δk and $k_F$ values

The emergent soft gap due to electron-electron interactions follows a DOS of $g(E) \propto |E - E_F|^{d-1}$. Since we have established a three-dimensional band structure, we are looking at a quadratic distribution of spectral weight around $E_F$. To be able to compare this relationship with experimental data, thermal and instrument broadening needs to be taken into account, as in $I(E) = DOS(E)f(E,T)\otimes R(E)$ where $f(E,T)$ is the Fermi-Dirac distribution and $R(E)$ represents instrumental energy resolution. Beamline 7.0.2 has a total resolution below 10meV, while the kBT at 100 K (200K) is 8.6 meV (17.23 meV). A correction to the experimental data I(E) can be done dividing by a Fermi-Dirac Function $f(E,T) = \frac{1}{1+exp\left(\frac{E-E_F}{k_B T}\right)}$.

The expected region of depleted DOS is very small for the x=0.1 and x=0.2 compositions. Additionally, broadening of the underlying $e_g$ bands may add background intensity and reduce the area where a quadratic reduction in DOS can be resolved. We therefore limit our quantitative study of the soft gap to x=0.8. For the x=0.8 case, power fits and residuals are shown in Figure S6.4. We note that extending the soft gap range is possible but with lower fit accuracy. We also present examples of fits starting at 0.065 eV instead of at $E_F$, since this allows for a longer fit. These results indicate that the 350 meV soft gap estimate is a lower bound for the soft gap. Small deviations and fit uncertainty in the DOS, $|E - E_F|^{pgap}$ from $p_{gap} = 2$ are shown in the inset.

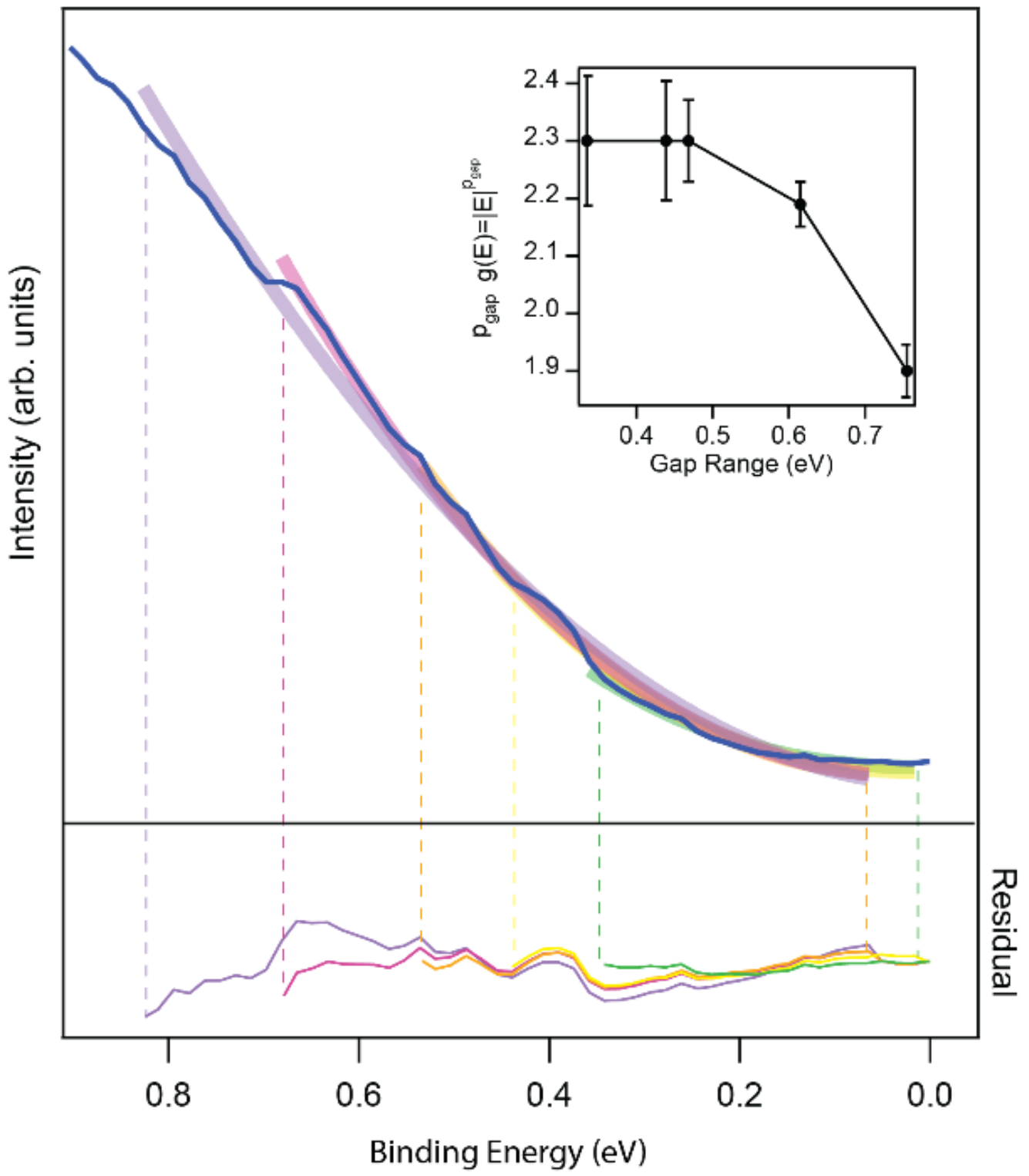

**Figure S6.4** Detailed study of the soft gap for the x=0.8 case. Several fits are shown with their residuals in each case. The inset shows the power fit exponent and fit uncertainty for each case, with the bottom axis indicating the fit range in eV.

Extended range spectra for integrated EDC across the whole Brillouin zone is shown below in Figure S6.5 a while temperature dependence of the low binding energy region is shown in Figure S6.5b. No evidence of coherent/incoherent substantial intensity ratio change was observed.

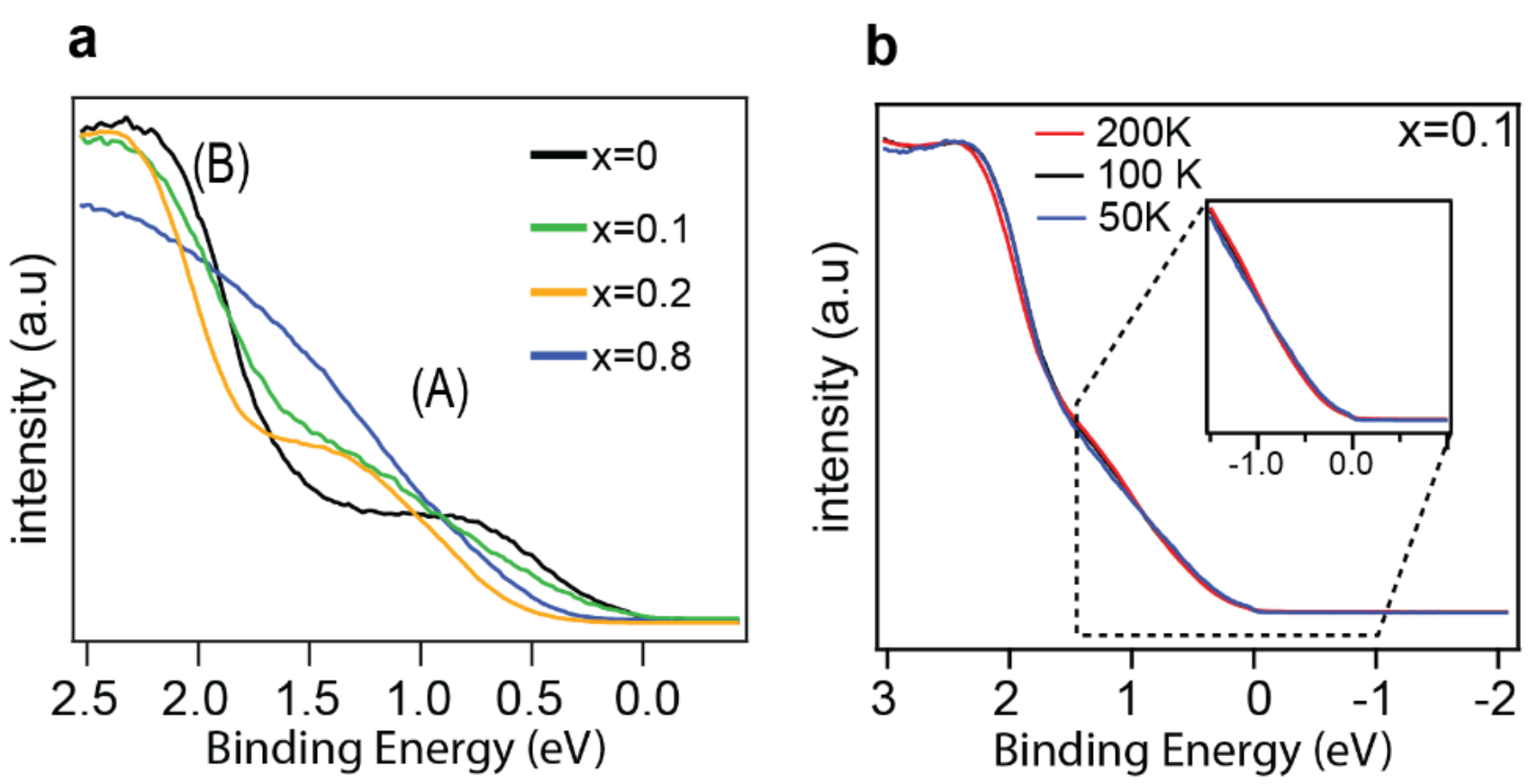

**Figure S6.5** (a)Integrated valence band for x=0, 0.1, 0.2 and 0.8 showcasing the spectral weight transfer and intensity loss at EF. Features (A) and (B) correspond to TM eg and TM $t_{2g}$ states respectively. (b) Temperature dependence of the valence band for x=0.1. An inset showing the $e_g$ region highlights the minimum differences observed.

## S7. Additional TEM and EDS data and discussion

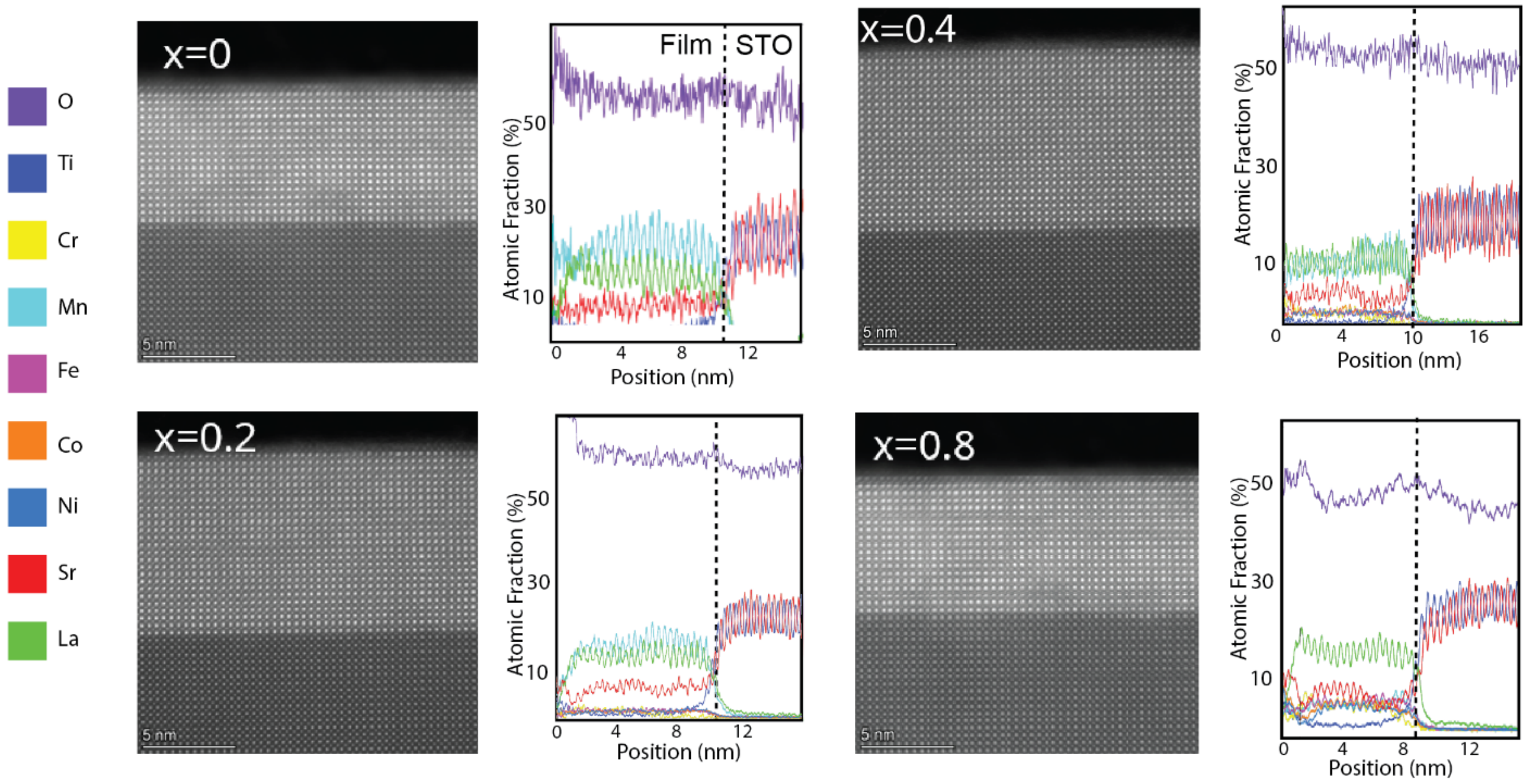

**Figure S7**. 1 HAADF and EDS profiles for compositions x=0, 0.2,0.4 and 0.8.

Cross-sectional HAADF-STEM shows epitaxial crystalline films in the regions examined, with no resolved low-angle grain boundaries or dislocations. EDS maps indicate substantial transition-metal intermixing, while the through-thickness profiles reveal surface/interface compositional variations, including Sr and Cr surface enrichment and Mn enrichment near the interface in some examined samples. The present measurements therefore support substantial intermixing. Surface/interface gradients may contribute an additional, spatially correlated component to the disorder landscape.